\documentclass{aa}  

\usepackage{graphicx}
\usepackage{xcolor}
\usepackage{txfonts}
\usepackage{hyperref} %[options]
\begin{document}

   \title{The chemical evolution of the Milky Way thin disk using solar twins}

   % \subtitle{I. Overviewing the $\kappa$-mechanism}

   \author{A. Plotnikova,
          \inst{1}
          L. Spina,
          \inst{2}
          B. Ratcliffe\inst{3},
          G. Casali\inst{4,5,6},
          \and
          G. Carraro
          \inst{1}
          }

   \institute{Dipartimento di Fisica e Astronomia, Universit\'a di Padova,
              Vicolo dell'Osservatorio 3, I-35122 Padova, Italy\\
              \email{anastasiia.plotnikova@studenti.unipd.it}
         \and
             INAF, Osservatorio Astronomico di Arcetri, Arcetri, Italy
         \and 
             Leibniz-Institut für Astrophysik Potsdam (AIP), An der Sternwarte 16, 14482 Potsdam, Germany
         \and 
             Research School of Astronomy \& Astrophysics, Australian National University, Cotter Rd., Weston, ACT 2611, Australia
         \and 
             ARC Centre of Excellence for All Sky Astrophysics in 3 Dimensions (ASTRO 3D), Stromlo, Australia
         \and 
             INAF-Osservatorio di Astrofisica e Scienza dello Spazio di Bologna, via P. Gobetti 93/3, 40129, Bologna, Italy
             }

   \date{Received , ; accepted , }

  \abstract    
   {}
   {In this study we address whether the age--metallicity relation (AMR) deviates from the expected trend of metallicity increasing smoothly with age. We also show the presence (or absence) of two populations, as recently claimed using a relatively small dataset. Moreover, we studied the Milky Way thin disk's chemical evolution using solar twins, including the effect of radial migration and accretion events.}
   {In particular, we exploited high-resolution spectroscopy of a large sample of solar twins in tandem with an accurate age determination to investigate the Milky Way thin disk age--metallicity relationship. Additionally, we derived the stars' birth radius and studied the chemical evolution of the thin disk.}
   {We discovered that statistical and selection biases can lead to a misinterpretation of the observational data. An accurate accounting of all the uncertainties led us to detect no separation in the AMR into different populations for solar twins around the Sun ($-0.3 < [Fe/H] < 0.3$ dex). This lead us to the conclusion that the thin disk was formed relatively smoothly. For the main scenario of the Milky Way thin disk formation, we suggest that the main mechanism for reaching today's chemical composition around the Sun is radial migration with the possible contribution of well-known accretion events such as \textit{Gaia}-Enceladus/Sausage (GES) and Sagittarius (Sgr).}
   {}

   \keywords{Milky Way disk -- solar twins --
                star's abundances --
                stellar ages
               }

  \titlerunning{Thin disk chemical evolution}
  \authorrunning{Plotnikova et al.}
   \maketitle

%-------------------------------------------------------------------

\section{Introduction}
The formation and evolution of the Milky Way are among the most important topics in modern astronomical research. Using the advantage of our location from inside the Milky Way we can study it in great detail. Based on different observational results, several models of Galaxy formation were suggested in the past. The first to form was the infant metal-poor bulge through rapid collapse  (see, e.g., \citet{Plotnikova_2023} and references therein). Then 11 - 12 Gyr ago the thick disk and the halo were formed \citep[][]{Chiappini1997} according to the simple closed box model with instantaneous mixing of produced elements \citep[][]{Pagel_1997} where metallicity increases smoothly with time \citep[][]{Sahlholdt_2022}. Later on, around 10 Gyr ago, the thin disk started to form \cite[][]{Chiappini1997}. Its formation created a double population in the alpha-metallicity diagram \citep[][]{Yoshii_1982, Gilmore_1983}. According to the  \citet[][]{Chiappini1997} model, two infall episodes originated the halo-thick disk and thin disk. About 10 Gyr is the time found for the second infall in the revised two-infall model by \citet{Spitoni_2019}, which was designed to reproduce the two alpha-abundance sequences observed in the solar neighborhood. More recently, several other models were suggested that assumed as second infall the accretion of the gas previously ejected into the halo \citep[][]{Khoperskov_2021} or accreted from some major merger events, such as the \textit{Gaia}-Enceladus/Sausage (GES) \citep[][]{Bignone_2019, Buck_2020}.

In this paper we focus on the thin disk formation as this work's main target of investigation, and we would like to address the following questions: What  the key moments in the evolution of our Galaxy are, and whether its most recent evolution has occurred in substantial isolation from the surrounding environment \citep[e.g.,][]{Snaith2015} or instead if external factors have strongly contributed to the creation of multiple stellar populations \citep[e.g.,][]{Chiappini1997}. The latest literature is rich in observational studies suggesting that the recent evolution of the Galaxy has been influenced by the interaction with dwarf galaxies \citep[e.g.,][]{Ruiz-Lara_2020, BlandHawthorn21, Lu_2022, Antoja_2022, Gondoin_2023, Ratcliffe_2023} merging with the Milky Way, such as the GES \citep{Belokurov2018,Helmi_2018} or Sagittarius (Sgr) \citep{Ibata_1994}. The same scenario is also supported by theoretical studies and simulations \citep[e.g.,][]{Laporte_2019,Wang_2024_Sgt}. 

Currently, the  different models have tiny differences between them and we need a more detailed analysis of high-quality observational data to be able to distinguish them. We are lucky to have access to several high-resolution spectroscopic datasets such as HARPS and APOGEE, and high-quality astrometry from \textit{Gaia} Data Release 3 (DR3). Altogether, they offer us a great opportunity to study the Milky Way formation history in great detail. One of these important details is the age-metallicity relation. A recent study by \citet{Nissen_2020} has found a bimodality in the age-metallicity diagram traced by nearby solar-twin stars. This result suggests the existence of two separate populations within the thin disk resulting from two episodes of accretion of gas onto the Galactic disk with a quenching of star formation around 5-6 Gyr ago. The study of \citet{Nissen_2020} was based on only 72 stars and, as stated in the paper, it should be proven with a bigger dataset to be a solid statistical result; however, their results seem to agree with more recent studies based on APOGEE DR17 data\footnote{\url{https://www.sdss4.org/dr17/}} \citep[][]{Jofre_2021, Anders_2023}. In particular, \citet{Jofre_2021}  shows a clear discontinuity in the [C/N]-metallicity diagram of red clump stars in the solar neighborhood. For red clump stars the [C/N] abundance ratio can be considered a proxy of stellar ages, and so the two populations found by \citet{Jofre_2021} can be traced back to those of \citet{Nissen_2020}. \citet{Anders_2023}  use chemical APOGEE abundances to derive the ages of red giant stars confirming the bimodality of the age-metallicity diagram suggested by \citet{Nissen_2020} (see Fig. 8 in \citealt{Anders_2023}) and \citet{Jofre_2021}. The result should not be surprising as their ages were mostly inferred by the C and N abundances, which are the same analyzed by \citet{Jofre_2021}. However, the actual presence of these populations in the thin disk is still a matter of debate. Several other studies show no signature of two populations \citep[][]{Xiang_2022, Lu_2022, Miglio_2021}. Interestingly, all of these studies use different age determination methods, targets from different observations, and cutoffs, which are the crucial points on this issue \citep{Sahlholdt_2022, Queiroz_2023}. 

According to the simple closed box model with instantaneous mixing of produced elements \citep{Pagel_1997}, the metallicity and abundances of other elements are expected to increase smoothly with time. However, it was found that the chemical trends of stars with different ages at the same location in the Galaxy intersect with each other. Important insights in this context were provided by \citet{Ratcliffe_2023} based on APOGEE data and age estimations from \citet{Anders_2023}. The authors compute the birth radii of red giant branch stars based on the assumption of the presence of radial migration, and use that information to study the chemical evolution at different locations across the Galactic disk. The chemical trends of the stars with different ages for the same birth radii are instead distributed smoothly, which means that radial migration can explain the chemically mixed composition in the Galactic thin disk stars. However, they reported the presence of fluctuations in the metallicity  and [X/H]-gradient evolution, which they tentatively ascribe to the dilution in [Fe/H] from gas brought by the GES and Sgr mergers. Two of these features are observed at lookback times of $\sim$4 and $\sim$6 Gyr. Altogether it produces a scenario of Milky Way formation with radial migration and accretion events playing an important role in the chemical composition of the thin disk.

For this work we focused on   thin disk solar twins. Due to their similarity to the Sun, we can perform high-accuracy differential spectroscopic analysis, which  makes them the perfect candidates to study chemical evolution in detail. We studied the age-metallicity relation (AMR) for the dataset of 485 stars with high-resolution spectroscopic observation. We studied the main sources of misinterpretation of the data, such as statistical and selection biases and different types of uncertainties. In addition, we derived the birth radii for stars under investigation to study the ability of the radial migration model to explain observed chemical trends. Moreover, we   searched for signatures of the most massive mergers.

In Sect. \ref{data} we explain how we obtained all the different stellar properties. In Sect. \ref{AMR} we talk about the method we used to study the AMR, and we present our results with a detailed analysis of all possible uncertainty sources. In Sect. \ref{Chemical trends} we study how the radial migration model together with the presence of major merger events can explain the observational features. We present our conclusions in Sect. \ref{conclusion}. 

\section{Data}\label{data}
In this work, we make use of two datasets of spectroscopic parameters and abundances of solar twins (i.e., effective temperature $T_{eff}: T_{eff,\,\,\,\odot} \pm 200$ K; surface gravity $logg:logg_{\odot} \pm 0.20$ dex; metallicity $[Fe/H]: [Fe/H]_{\odot} \pm 0.3$ dex; where $T_{eff,\,\,\,\odot} = 5771$ K, $logg_{\odot} = 4.44$ dex and $[Fe/H]_{\odot} = 0.0$ dex; \citealt{Ayres_2006}) with high-resolution high signal-to-noise spectroscopic observation to have a good number of stars for the statistics. The first dataset consists of HARPS-North (HARPS-N) spectra for 114 stars, which were fully reduced and analyzed for this work, as described in  Sect. \ref{sec:spectroscopy}. The outcomes of this spectroscopic analysis are the stellar parameters and chemical abundances. This dataset was greatly expanded by the chemical abundances derived by \citet{Casali_2020} for 371 solar-twin stars observed with the HARPS-South (HARPS-S) spectrograph. The ages and birth radii were consistently derived in this work for all stars in the two datasets (Sects. \ref{age determination} and \ref{sec: Rbirth}).

\subsection{Spectroscopic analysis}\label{sec:spectroscopy}

\subsubsection{Data sample and data reduction}
To obtain high-quality spectra we used the observations available in the archive of the  HARPS-N spectrograph. This  is a high-precision spectrograph mounted on the \textit{Galileo} National Telescope (TNG) in La Palma Island (Canary Islands, Spain). With HARPS-N we are able to obtain high-resolution (R=115,000) optical spectra with a broad wavelength coverage (378-691 nm).

In the archive IA2\footnote{\url{http://archives.ia2.inaf.it/tng/}} we queried for stars with \texttt{CCF Mask} equal to G2 which should correspond to G-type stars. We obtained reduced spectra for 1118 stars and most of them had several exposures. Among these, we only used spectra with $S/N>30 px^{-1}$. 

All spectra were normalized by the IRAF\footnote{\url{https://iraf-community.github.io/pyraf.html}} function \texttt{continuum} and were Doppler-shifted by \texttt{dopcor} using spectral radial velocities derived by \texttt{crosscorrRV}\footnote{\url{https://pyastronomy.readthedocs.io/en/latest/pyaslDoc/aslDoc/crosscorr.html}} through cross-correlation with the solar spectrum. Then, all exposures were stacked together by a Python script that re-bins each spectrum to the common wavelengths without changing the resolution, computes the median of the pixels, and applies $3\sigma$ clipping to the pixel values. After the spectra were combined, we were able to achieve for some stars the signal-to-noise ratio $2500 px^{-1}$. During the analysis, all spectroscopic binaries were removed together with misclassified solar-type stars based on cross-correlation analysis with respect to the solar spectrum.

Additionally, we analyzed solar spectra obtained through observations of asteroids, planets, and the planets' satellites  (i.e., Vesta, Venus, Europa, Ganymede). As detailed in the following paragraphs, the solar spectrum was essential to perform a line-by-line differential analysis of our stellar sample with respect to the Sun.

\subsubsection{Stellar parameters and chemical abundances}
The method of spectroscopic analysis applied in this work is the line-by-line differential analysis relative to the solar spectrum, which is also the same method applied by \citet{Casali_2020} to derive their chemical abundances. This method is perfectly suited for the analysis of solar-twin stars as it cancels out the impact of log~gf parameters in the error balance and also reduces the impact of systematics in models and in the spectrum normalization. As a result, the line-by-line differential analysis lets us derive stellar parameters and chemical abundances with extremely high precision (\citet{Ramirez_2014}, \citet{Melendez_2014}, \citet{Spina_2016}, \citet{Spina_2018}, \citet{Nissen_2020}, \citet{Casali_2020}). In our analysis, we also adopted the same code and linelists used by \citet{Casali_2020}. To derive chemical abundances of 25 chemical elements (C, Na, Mg, Al, Si, S, Ca, Sc, Ti, V, Cr, Mn, Co, Ni, Cu, Zn, Sr, Y, Zr, Ba, La, Ce, Nd, Sm, Eu) we used the master list of atomic transitions of \citet{Melendez_2014} that includes 98 lines of Fe I, 17 of Fe II, and 183 for the other elements, detectable in the HARPS spectral range ($3780-6910$ \AA), and measured their equivalent widths (EWs) with \texttt{Stellar diff}.\footnote{\url{https://github.com/andycasey/stellardiff}}

The \texttt{Stellar diff} code allows the user to interactively select one or more spectral windows to create a mask around each studied line that contains the measuring line and parts of the local continuum around it without any contamination from other lines. Choosing the continuum locally allows us to minimize the uncertainties due to imperfect spectrum normalization and unresolved features in the continuum (\citet{Bedell_2014}). The code fits each line of interest with a Gaussian profile and provides the EW and uncertainty. Additionally, \texttt{Stellar diff} is able to identify hot pixels and cosmic rays and remove them from the analysis. 

\begin{figure}
\centering
\includegraphics[scale=0.5]{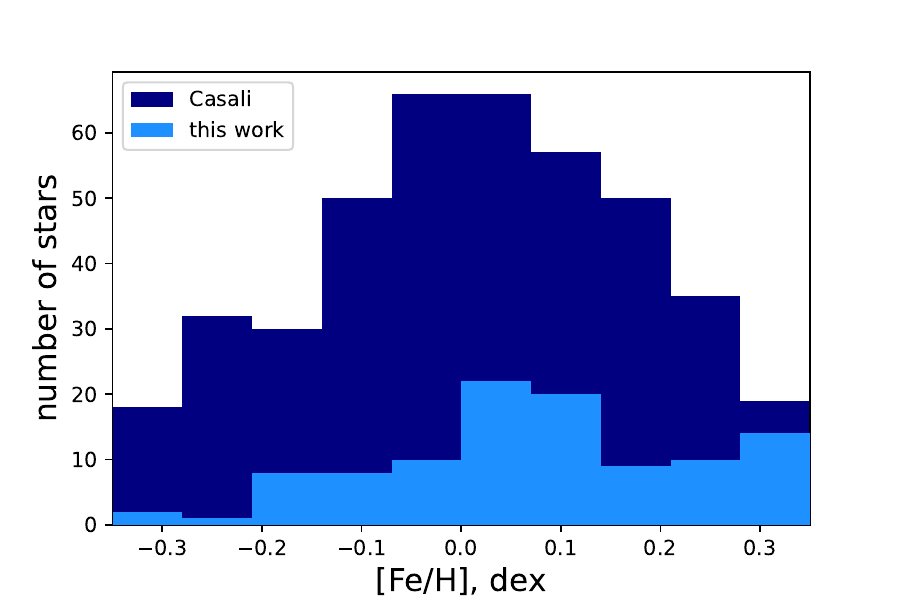}
\caption{Metallicity distribution for the dataset analyzed in this work and for the dataset from \citet{Casali_2020}.}
\label{fig:metallicity distribution for Casali and ours}
\end{figure}

\begin{figure}
\centering
\includegraphics[scale=0.46]{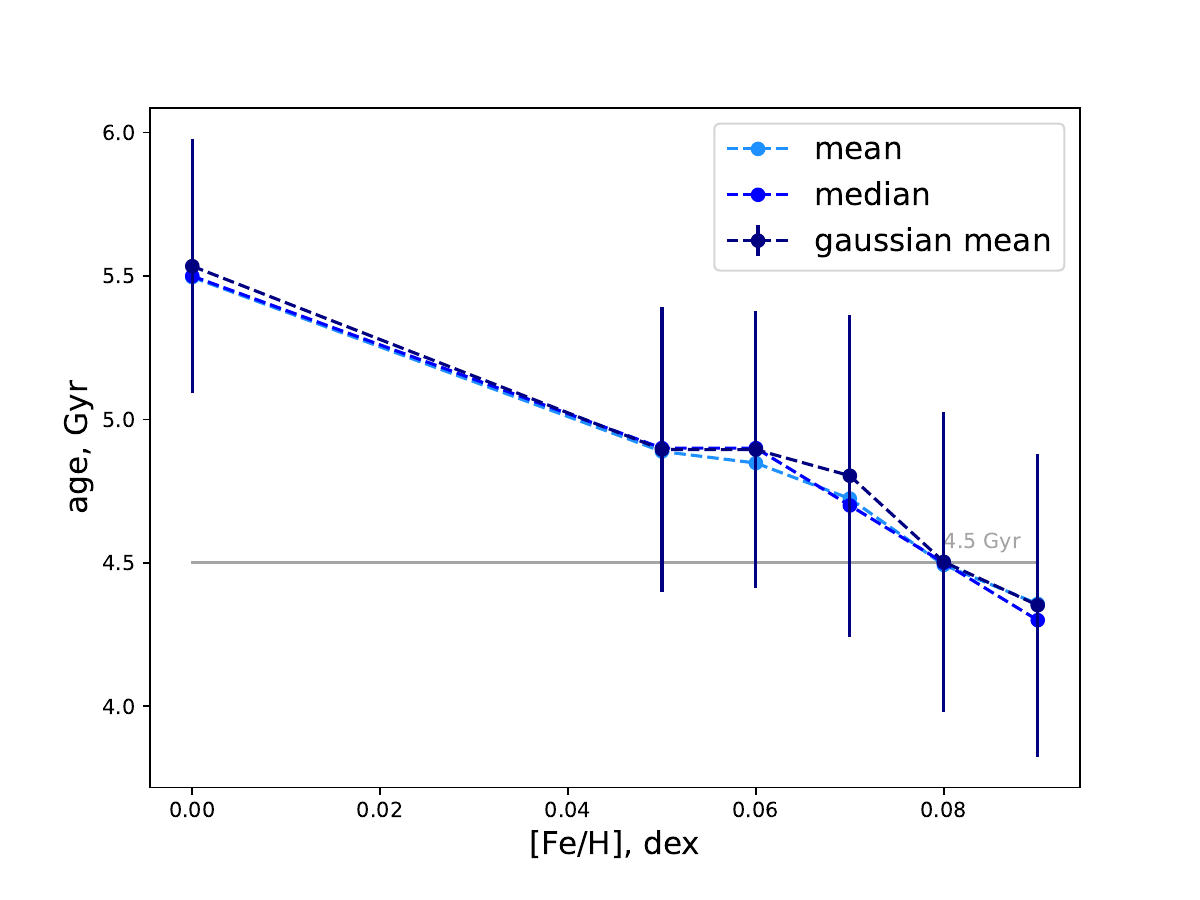}
\caption{Age of the Sun calculated for different values of metallicity to calibrate isochrones for the influence of atomic diffusion at the solar age. Light blue represents the mean values for the Sun's age, blue is the  median, and dark blue is the Gaussian mean. The gray line is the most accurate estimate of the Sun's age \citet{Connelly_2008, Amelin_2010}.}
\label{fig:age calibration with the Sun}
\end{figure}

To determine the stellar parameters we used the \texttt{qoyllur-quipu (q2)}\footnote{\url{https://github.com/astroChasqui/q2}} code (\citet{Ramirez_2014}). This code performs line-by-line differential analysis of the  EWs of the iron lines relative to the measurements of the Sun. The \texttt{q2} algorithm iteratively searches for three equilibria: extinction, ionization, and the trend between the iron abundances and the reduced EW $log[EW/\lambda]$. Iterations start with the initial parameters (normal solar parameters) and produce a final set of parameters that satisfy all three equilibria. For the analysis we used the  Kurucz (ATLAS9) grid of model atmospheres \citep{Castelli_2004}, the MOOG 2014 version \citep{Sneden_1973}, and the following solar parameters: $T_{eff}=5771$ K, $log g = 4.44$ dex, $[Fe/H] = 0.00$ dex, and $\xi = 1.00$ km/s \citep{Ayres_2006}. The uncertainty evaluation procedure is described in \citet{Epstein_2010} and \citet{Bensby_2014} where the relation between stellar parameters is taken into account. The typical uncertainty for each parameter, which is the average uncertainty of all stars, is $\sigma(T_{eff}) = 9$ K, $\sigma(log g) = 0.03$ dex, $\sigma([Fe/H]) = 0.007$ dex, and $\sigma(\xi) = 0.02$ km/s. In Fig. \ref{fig:metallicity distribution for Casali and ours} we present the metallicity distribution for the dataset analyzed in this work and for the dataset from \citet{Casali_2020}. We can see that both datasets are consistent with each other.

%%%%%% place for metallicity distribution plot

From stellar parameters and their uncertainties \texttt{q2} using appropriate atmospheric models derives chemical abundances for each of the  25 chemical elements. All the abundances are scaled relative to the Sun's measurements on a line-by-line basis. In addition, with the  \texttt{blends} driver in the MOOG code, \texttt{q2} takes into account the hyper-fine splitting (HFS) effects in Y, Ba, and Eu. The HFS line list is adopted from \citet{Melendez_2014}. However, this approach still leaves the possibility for additional uncertainty. Ideally, each line should be fully modeled and the observed shape should be compared with the modeled one \citep[][]{Bensby_2005, Feltzing_2007}, although our analysis is robust enough for stars with close-to-solar stellar parameters. Finally, \texttt{q2} derives uncertainties for every abundance [X/H] by the quadratic sum of the line-by-line scatter due to EW measurements (standard error) and errors of the atmospheric parameters. For the chemical elements with just one line measurement (Sr, Eu) as a standard error, we took the uncertainty of the EW measurement with \texttt{Stellar diff}.

\subsubsection{Comparison of two datasets}
The spectra for the dataset of solar twins analyzed in this work (HARPS-N) and dataset from \citet{Casali_2020} (HARPS-S) were acquired by spectrographs with similar characteristics. The same spectral analysis was aslo applied to both of these datasets. However, there may   still be some small systematics in abundance determinations of the two datasets, due to minor differences between the instruments.

There are 27 stars in common between \citet{Casali_2020} and our dataset. This allowed us to compare the precision and accuracy of the abundances and stellar parameters derived by us and by \citet{Casali_2020}. To this end, we performed a linear regression ($y=kx+b$) in [X/H]$_{this\,\,\,work}$ versus [X/H]$_{this\,\,\,work}-$[X/H]$_{Casali}$ space. As a result, for most of the chemical elements we do not see the correlation between the abundance difference and the abundance value derived in this work (Table \ref{tab:abundances shift}, the slope consistent with zero within 2$\sigma$ or zero (Ca, Na)). We only see a constant systematic shift between \citet{Casali_2020} and our sample abundances. Only a few elements show a slightly increasing difference between the two datasets with increasing abundance value (Table \ref{tab:abundances shift}: Mg, Sc, Zn, V), but they are only 20\% of the whole amount of elements. Therefore, we only applied the shift correction for \citet{Casali_2020} data to bring all abundances to the same point (Table \ref{tab:abundances shift}). This correction was computed as a median shift between these two abundances for each element.

\begin{table}[]
    \centering
    \begin{tabular}{lccccc}
        % \hline
        \hline
        element&k&$\Delta$k&b&$\Delta$b&shift\\
        \hline
        CI & 0.000 & 0.025 & 0.044 & 0.009 & 0.059\\
        NaI & 0.000 & 0.029 & 0.019 & 0.005 & 0.029\\
        MgI & 0.124 & 0.051 & 0.043 & 0.007 & 0.067\\
        AlI & 0.012 & 0.039 & 0.015 & 0.002 & 0.042\\
        SiI & 0.023 & 0.034 & 0.023 & 0.004 & 0.034\\
        SI & 0.033 & 0.035 & 0.045 & 0.010 & 0.031\\
        CaI & 0.045 & 0.034 & 0.029 & 0.005 & 0.035\\
        ScI & 0.081 & 0.046 & 0.032 & 0.008 & 0.060\\
        TiI & 0.030 & 0.030 & 0.024 & 0.004 & 0.031\\
        VI & 0.061 & 0.034 & 0.031 & 0.007 & 0.024\\
        CrI & 0.012 & 0.042 & 0.029 & 0.005 & 0.039\\
        MnI & -0.026 & 0.027 & 0.030 & 0.006 & 0.027\\
        CoI & 0.046 & 0.040 & 0.034 & 0.007 & 0.033\\
        NiI & 0.005 & 0.030 & 0.021 & 0.005 & 0.027\\
        CuI & -0.034 & 0.041 & 0.046 & 0.009 & 0.041\\
        ZnI & 0.064 & 0.040 & 0.020 & 0.005 & 0.038\\
        SrI & 0.007 & 0.001 & 0.009 & 0.002 & 0.001\\
        ScII & 0.022 & 0.044 & 0.034 & 0.007 & 0.042\\
        TiII & 0.041 & 0.042 & 0.030 & 0.005 & 0.041\\
        CrII & 0.011 & 0.033 & 0.021 & 0.004 & 0.029\\
        % YII & 0.041 & 0.006 & 0.023 & 0.004 & 0.002\\
        % ZrII & 0.155 & 0.032 & 0.060 & 0.009 & 0.042\\
        % BaII & 0.062 & 0.023 & 0.027 & 0.005 & 0.022\\
        % CeII & 0.413 & 0.015 & 0.080 & 0.014 & 0.016\\
        % NdII & 0.318 & 0.015 & 0.044 & 0.007 & 0.029\\
        % SmII & 0.099 & -0.036 & 0.133 & 0.018 & -0.048\\
        % EuII & 0.443 & 0.023 & 0.150 & 0.022 & 0.085\\
        \hline
    \end{tabular}
    \caption{Comparison of abundances derived in this work and abundances derived by \citet{Casali_2020} for the 27 stars in the overlap.}
    \label{tab:abundances shift}
\end{table}

\begin{figure*}
\centering
\includegraphics[scale=0.5]{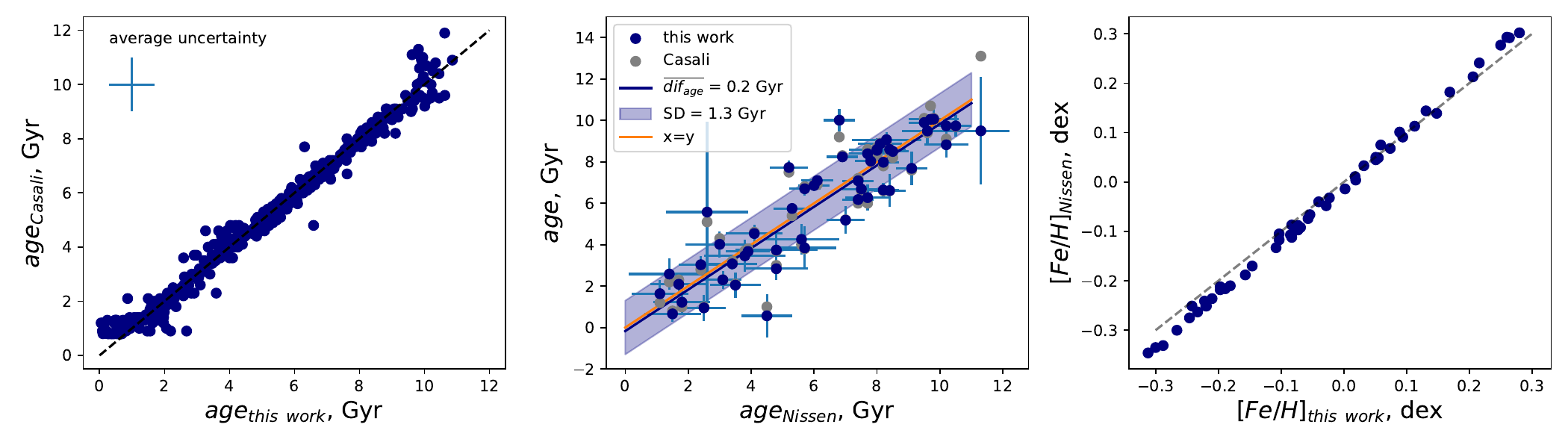}
\caption{Comparison of age and metallicity derived by different methods. Left: Comparison of age derived in this work and in \citet{Casali_2020}. Middle: Comparison of age derived in this work and in \citet{Nissen_2020} (blue circles), its mean difference ($\overline{dif_{age}}$), and standard deviation (SD); comparison of age derived in \citet{Casali_2020} and in \citet{Nissen_2020} (gray circles). Right: Comparison of metallicity derived in this work and in \citet{Nissen_2020}.}
\label{fig:age comparison}
\end{figure*}

\subsection{Distance determination}\label{Distance determination}
Distance is one of the most important parameters in dynamics studies. \citet{Plotnikova_2022} showed that, for the bright stars in the solar vicinity, the directly inverted \textit{Gaia} Early Data Release 3 (EDR3) parallaxes are the best distance estimation method. Since most of the stars under consideration are bright and are located inside 200 pc around the Sun, for distance determination we used only \textit{Gaia} EDR3 parallaxes (\citet{GaiaEDR32021}).
\subsection{Astrometric parameters}
Intending to calculate orbits and orbital parameters, we extracted the coordinates and proper motion components from the \textit{Gaia} DR3 archive \citep[][]{Gaia_Collaboration_2022_DR3}. The uncertainties for the astrometric parameters are shown in Table \ref{tab:Unc GEDR3 astrometry}. Radial velocities were obtained from the spectroscopic analysis. The position and velocity of the stars in  equatorial coordinates were transformed to galactocentric coordinates by the  \texttt{astropy}\footnote{\url{https://docs.astropy.org/en/stable/_modules/astropy/coordinates/builtin_frames/galactocentric.html astropy}} Python package.

\begin{table}[]
    \centering
    \begin{tabular}{lccc}
        % \hline
        \hline
        Data product or source type&\multicolumn{3}{c}{Typical uncertainty}\\
        % \cline{2-5}
        &G $<$ 15&G = 17 &G = 20 \\
        \hline
        Five-parameter astrometry&&&\\
        \hline
        position, mas&0.01 - 0.02&0.05&0.4\\
        parallax, mas&0.02 -
        0.03&0.07&0.5\\
        proper motion, mas yr$^{-1}$&0.02 - 0.03&0.07&0.5\\
        \hline
        Six-parameter astrometry&&&\\
        \hline
        position, mas&0.02 - 0.03&0.08&0.4\\
        parallax, mas&0.02 - 0.04&0.1&0.5\\
        proper motion, mas yr$^{-1}$ & 0.02 - 0.04&0.1&0.6\\
        \hline
    \end{tabular}
    \caption{Uncertainties of \textit{Gaia} Early Data Release 3 astrometry \citet{GaiaEDR32021}}
    \label{tab:Unc GEDR3 astrometry}
\end{table}

\subsection{Kinematics}
We obtained orbits and orbital parameters for all stars in the dataset by numerical calculation and adopting a model of Galactic axisymmetric potential. We corrected the velocities of the stars by the velocity of the Sun with respect to the Galactic center, which is computed independently from the velocity of the local standard of rest \citep[][]{Reid_2004_vSun, GRAVITY_Collaboration_2018, Drimmel_2018_vSun}:

\begin{equation}
    v_{R\odot} = -12.9 \pm 3.0 \text{ km s}^{-1}
,\end{equation}
\begin{equation}
    v_{\phi\odot} = 245.6 \pm 1.4 \text{ km s}^{-1}
,\end{equation}
\begin{equation}
    v_{Z\odot} = 7.78 \pm 0.09 \text{ km s}^{-1}
.\end{equation}
Here the distance between the Galactic center and the Sun is $R_0 = 8.122 \pm 0.033$ kpc (\citet{GRAVITY_Collaboration_2018}).

We used the \citet{McMillan_2017} Galactic potential without bar implementation in \texttt{galpy}\footnote{\url{http://github.com/jobovy/galpy}} \citep[][]{Bovy_2015_galpy} to calculate the orbits and orbital parameters, such as the total energy ($E_{n}$), eccentricity ($e$), apo-center ($R_{apo}$), peri-center ($R_{peri}$), guiding radius ($R_{guide}$), angular momentum along the z-axis ($L_z$), and the velocity components in different directions ($v_r$, $v_R$, $v_{\phi}$, $v_z$).

\subsection{Age determination}\label{age determination}
We derived ages using the isochrone fitting technique on the Kiel diagram. To obtain the ages and their uncertainties, we made use of the algorithm developed by \citet{Plotnikova_2022}. Specifically, the parameters for each star were randomly independently selected 10,000 times from a Gaussian distribution, with mean as the parameter value and sigma as its uncertainty. For each selected set of parameters (out of 10,000), an age was computed. Afterward, the obtained age distribution was fitted with the Gaussian distribution; as a result, we take the mean as the age estimation and the sigma as the age uncertainty. The age fitting was carried out using the Padova isochrones PARSEC\footnote{\url{http://stev.oapd.inaf.it/cmd}} with the steps $\Delta age = 0.1$ Gyr, $\Delta [Fe/H] = 0.05$ dex; and for very young stars (initial guessed age is smaller than 1 Gyr) we repeated the algorithm with the steps $\Delta age = 0.01$ Gyr, $\Delta [Fe/H] = 0.05$ dex. Since we have effective temperature and surface gravity from high-precision spectroscopic analysis, we derived ages with a median uncertainty of 0.7 Gyr. It is also worth mentioning that it is necessary to calibrate these ages as a function of metallicity to take into account the influence of atomic diffusion at the solar age \citep[][]{Melendez_2012, Dotter_2017} and to obtain the last most accurate estimate for the age of the Sun \citep{Connelly_2008, Amelin_2010}. Therefore, we computed the age of the Sun for six different values of metallicity ([Fe/H]): 0.00, 0.05, 0.06, 0.07, 0.08, 0.09 dex. As   Fig. \ref{fig:age calibration with the Sun} shows, the metallicity equal to 0.08 dex shows the best agreement with the \citet{Connelly_2008} and \citet{Amelin_2010} solar age (4.5 Gyr). Therefore, we applied a 0.08 dex shift as a metallicity calibration.

\begin{figure}
\centering
\includegraphics[scale=0.5]{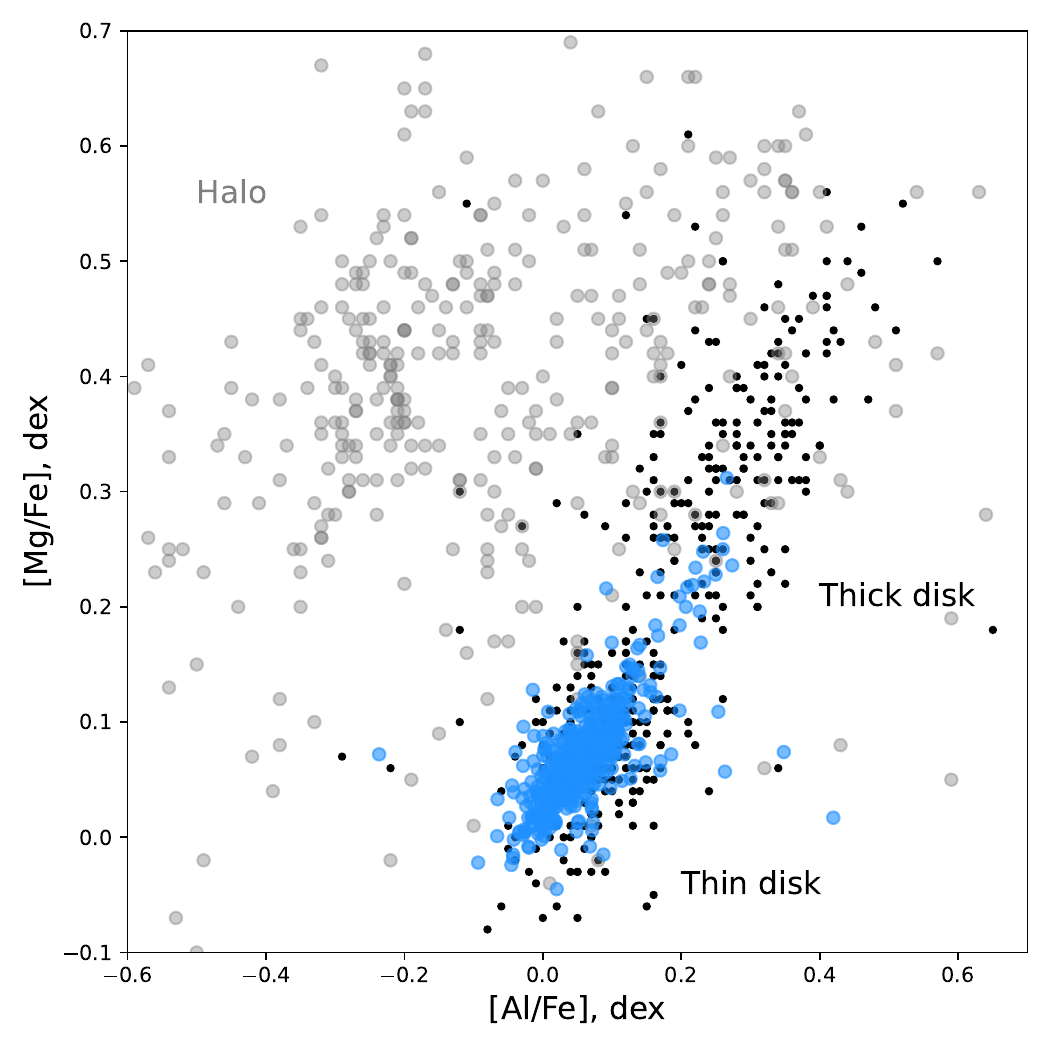}
\caption{Al-Mg correlation. The background: disk stars (black dots; \cite{Fulbright_2000}, \cite{Reddy_2003}, \cite{Simmerer_2004}, \cite{Reddy_2006}, \cite{Francois_2007}, \cite{Johnson_2012}, \cite{Johnson_2014}), halo stars (gray points; \cite{Yong_2013}), \cite{Roederer_2014}, and  dataset  (light blue circles; this work).}
\label{fig:AlMg}
\end{figure}

To check the accuracy of the determined ages we compared our results with the ages from \citet{Casali_2020} for 371 stars we have in common and the ages from \citet{Nissen_2020} for 52 stars. Figure \ref{fig:age comparison}-left shows excellent agreement between our age determinations and those from \citet{Casali_2020}. This is expected since both studies use a similar method of analysis. Specifically, both studies derived ages by isochrone fitting technique in surface gravity versus effective temperature space. Both studies also accounted for the atomic diffusion at the solar age.

Regarding the comparison between our age determinations to those from \citet{Nissen_2020}, it is worth mentioning that in both cases the ages were derived through isochrone fitting techniques, but in different parameter spaces.  \citet{Nissen_2020} used the $T_{eff}$ versus luminosity ($L$) space, while for this work we used   $T_{eff}$ versus $logg$. In Fig. \ref{fig:age comparison}-middle the average uncertainty for the two datasets is equal to 0.7 Gyr. The mean difference ($dif_{age}$) between the two measurements for the overlapping stars and its standard deviation (direct error) is

\begin{equation}
    \overline{dif_{age}} = \frac{\Sigma_{i=0}^{N} (age_{Nissen}^i - age_{this\,\,\,work}^i)}{N} = 0.17 \pm 1.30 \text{ Gyr}
,\end{equation}and the indirect error of the average difference is

\begin{equation*}
    \Delta\overline{dif_{age}} = \sqrt{\Delta\overline{age_{Nissen}}^2 + \Delta\overline{age_{this\,\,\,work}}^2} = 
\end{equation*}

\begin{equation}
    = \sqrt{0.7^2 + 0.7^2} = 0.9 \text{ Gyr}
.\end{equation}

The fact that direct (1.3 Gyr) and indirect (0.9 Gyr) errors are close to each other means that the ages from the two different methods are in good agreement. Additionally, the ages from \citet{Casali_2020} for these 52 stars show the same correlation as values from this work (Fig. \ref{fig: AMR}, middle). The tests with the \citet{Casali_2020} and \citet{Nissen_2020} ages both show that our age determination method is accurate and gives the same precision as in \citet{Casali_2020} and in \citet{Nissen_2020}.

\begin{figure*}
\centering
\includegraphics[scale=0.5]{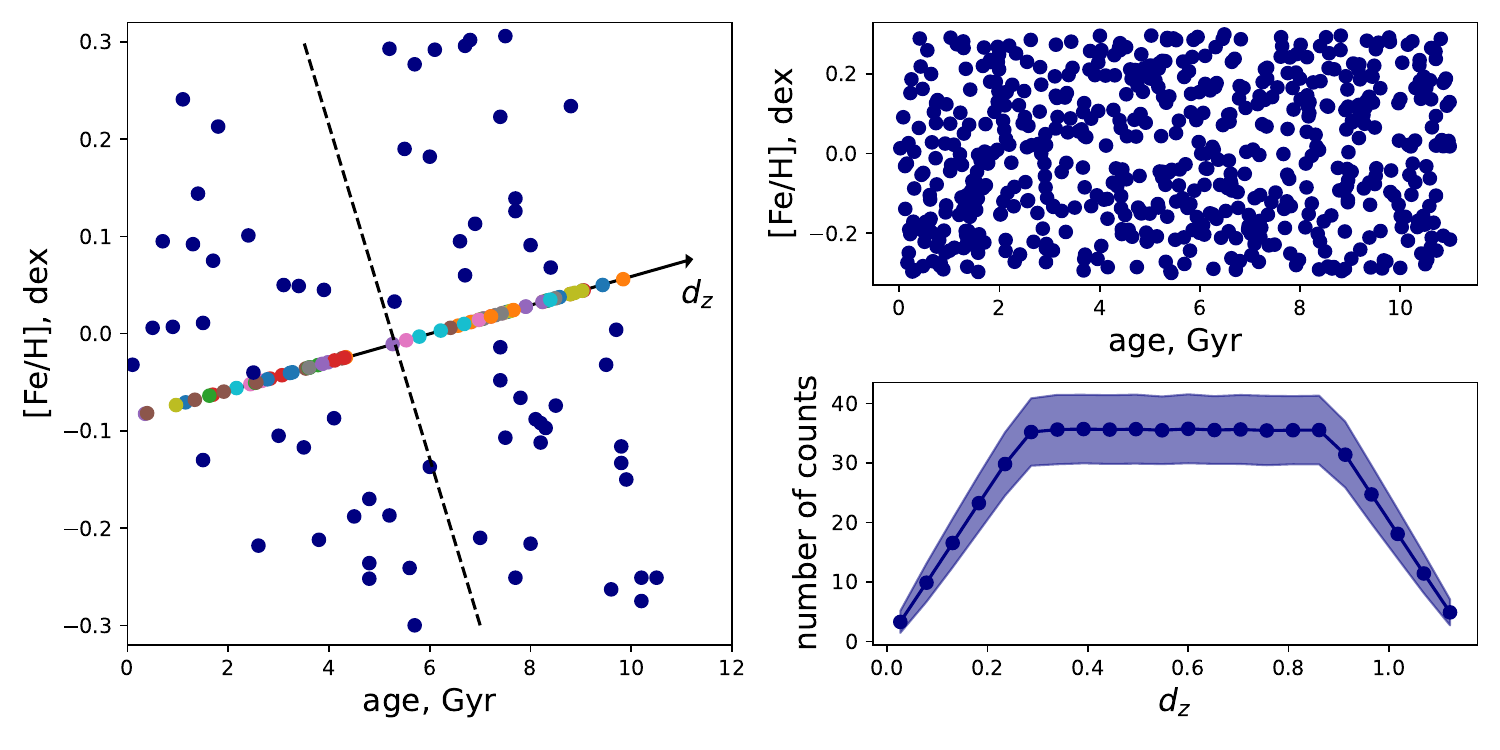}
\caption{Illustration of method used to test the separation in the age-metallicity map. Left: Age-metallicity map of Nissen's data (blue dots) with the division line for two populations (black dashed line) and with $d_z$ parameter axes (black solid line). The solar twins projected across the $d_z$ axis (colored points). Upper right: Random distribution of 600  stars. Lower right: Projection of the data from the upper right age-metallicity map to the $d_z$ axis.}
\label{fig: Method}
\end{figure*}

\subsection{Determination of the $R_{birth}$}\label{sec: Rbirth}
Chemical abundances contain information regarding the birth environment of the star \citep[][]{Freeman_2002, Ratcliffe_2022}, and are representative of when ($age$) and where ($R_{birth}$) the star was born \citep[e.g.,][]{Minchev_2018, Frankel_2018, Ness_2019, Lu_2022}. Since we have age and  metallicity, we can derive the birth radius using the method described in \citet{Lu_2022} and \citet{Ratcliffe_2023}, which has been shown to work in galaxies with stronger bars, for example the Milky Way (\citet{Ratcliffe_2024}). We assume that the star-forming gas in the Milky Way is azimuthally chemically homogeneous (observation: \citet{Deharveng_2000}, \citet{Esteban_2017}, \citet{Arellano-Cordova_2021}; simulations: \citet{Vincenzo_2018}, \citet{Lu_2022_sim}) and that the birth metallicity gradient is always linear in radius \citep[][]{Arellano-Cordova_2021, Esteban_2022}. Therefore, for any lookback time ($\tau$), we can write metallicity ($[Fe/H](R,\tau)$) as a function of the metallicity gradient at that time ($\nabla[Fe/H](\tau)$), birth radius, and   metallicity at the Galactic center ($[Fe/H](0,\tau)$):

\begin{equation}\label{Rbirth}
    [Fe/H](R_{birth},\tau) = \nabla[Fe/H](\tau) * R_{birth} + [Fe/H](0,\tau)
.\end{equation}By rearranging   Eq. \ref{Rbirth} we can estimate the birth radius as a function of age and metallicity:

\begin{equation}
    R_{birth}(age, [Fe/H]) = \frac{[Fe/H] - [Fe/H](0,\tau)}{\nabla[Fe/H](\tau)}
.\end{equation}The metallicity at the Galactic center and the metallicity gradient as a function of lookback time were taken from \citet{Ratcliffe_2023}.

\subsection{Origin}
To derive the origin of the dataset under investigation we used the Al-Mg correlation map, which    shows one of the best separations between halo, thin disk, and thick disk stars. As we can see in Fig. \ref{fig:AlMg}, almost all of our stars are lying in the location of the thin disk stars. By also taking into account that their stellar parameters are Sun-like, we consider them to be a good sample of thin disk stars.

\section{Age-metallicity relation}\label{AMR}

\subsection{Method}\label{method}
To study the presence of the two populations found by \citet{Nissen_2020} we used the following method. First, we determined the axis $d_z$ in the age-metallicity plot that maximizes the separation between the two populations. That axis was determined through the linear discriminant analysis\footnote{\href{https://scikit-learn.org/stable/modules/generated/sklearn.discriminant_analysis.LinearDiscriminantAnalysis.html\#}{sklearn.discriminant\_analysis.LinearDiscriminantAnalysis}} algorithm trained over the ages and metallicities used by \citet{Nissen_2020} (see their Fig. 3). Then, we rescaled the data (0.6 dex in metallicity to 12 Gyr in age) to let age and metallicity affect the separation equally. Finally, we projected all data points under investigation on $d_z$ and we studied the distribution of stars across that new axis. This procedure is illustrated in Fig. \ref{fig: Method}-left panel, where we show the age-metallicity diagram with the 72 solar twins by \citet{Nissen_2020}, the $d_z$ axis, the corresponding division line for the two populations, and the solar twins projected across the $d_z$ axis. 

While studying the distribution of stars across $d_z$ one should also consider the geometrical effects of that projection due to the edges of the dataset under consideration. As an illustrative example, the distribution of homogeneously distributed stars inside the same age and metallicity range covered by the \citet{Nissen_2020} sample (age: 0 - 12 Gyr, [Fe/H]: -0.3 - 0.3 dex) is shown in Fig. \ref{fig: Method}, right top (see all cutoffs in Table \ref{tab:cutoffs}). In Fig. \ref{fig: Method}, right bottom, the shape of the $d_z$ parameter distribution is not linear because of the geometry of the dataset. Instead, the distribution has two decreasing wings on the edges. All of this  occurs because the dataset lies in a rectangular space (i.e., there is no apparent correlation between ages and [Fe/H] abundances) and it is projected to the axis that is not parallel to one of the sides.

\begin{table}[]
    \centering
    \begin{tabular}{ll}
        % \hline
        \hline
        Criteria & range\\
        \hline
        CCF mask & G2\\
        spectroscopic binaries & removed \\
        misclassified by solar type stars & removed \\
        signal-to-noise, S/N & $>30 px^{-1}$\\
        effective temperature, $T_{eff}$ & $T_{eff, \odot} \pm 200$ K\\
        surface gravity, logg & $logg_{\odot} \pm 0.2$ dex\\
        metallicity, [Fe/H] & -0.3 - 0.3 dex\\
        age & 0 - 12 Gyr\\
        \hline
    \end{tabular}
    \caption{All cutoffs applied for the dataset under investigation. ($T_{eff, \odot} = 5771$ K, $logg_{\odot} = 4.44$ dex, \citet{Ayres_2006})}
    \label{tab:cutoffs}
\end{table}

Once we have determined the distribution of stars across the $d_z$ axis, we are interested in deriving the uncertainties in the star counting at each bin of the $d_z$ parameter. There are two types of uncertainties:
\begin{itemize}
    \item [1.] Measurement uncertainty associated with age and metallicity uncertainties of each star. We randomly and independently selected the metallicity and age values  1000 times
 from the Gaussian distribution with the mean as the measured value and with sigma as its uncertainty.
    \item [2.] Statistical uncertainty associated with bin counts,
    \begin{equation}
    \sqrt{\frac{1}{n_{bins}}(1-\frac{1}{n_{bins}})*n_{stars}}
    ,\end{equation}
    where $\frac{1}{n_{bins}}$ is the probability that the star falling into a chosen bin in $d_z$, and $n_{stars}$ is the number of stars in the dataset.
\end{itemize}

The method described above rigorously considers all the sources of uncertainty associated with the distribution of stars across the $d_z$ axis. Through this approach, we can establish whether or not there is significant evidence of the two populations identified by eye by \citet{Nissen_2020}. In Sect. \ref{results} we discuss how we applied this method to different datasets, including the one analyzed by \citet{Nissen_2020} and the much larger dataset provided in this study.

\begin{figure*}
\centering
\includegraphics[scale=0.5]{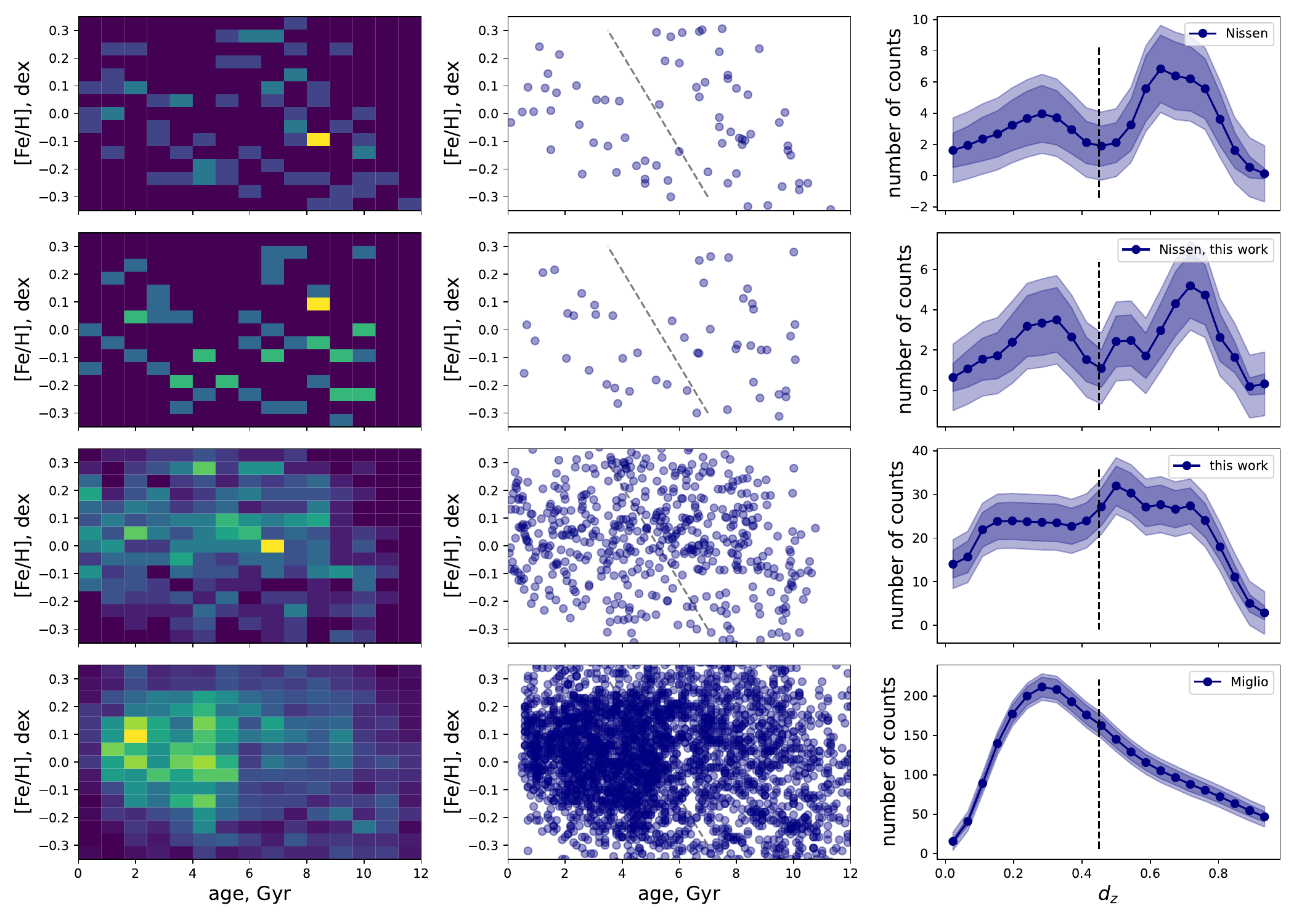}
\caption{AMR for three different datasets: [1] Nissen's dataset (72 stars, top row), [2] stars that we analyzed from Nissen's dataset (52 stars, second row), [3] dataset under investigation in this work (485 stars, third row), [4] dataset from \citet{Miglio_2021} (2785 stars, bottom row). The right column represents the $d_z$ separation parameter (Sect. \ref{method}) where the vertical dashed line represents the location of the Nissen's data drop, the dark blue region is the measurement uncertainty, and the light blue region is the quadratic sum of measurement and statistical uncertainty.}
\label{fig: AMR}
\end{figure*}

\subsection{Results}\label{results}
Now we apply the method explained in Sect. \ref{method} to study the presence of separation in the age-metallicity diagram pointed out by \citet{Nissen_2020}. It is important to mention that their dataset had only 72 stars, and \citet{Nissen_2020} claim their result should be checked with a bigger dataset. As our dataset includes 485 stars, it offers the possibility to further investigate this area in age-metallicity relation. In Fig. \ref{fig: AMR} we show the age-metallicity relations for four different datasets: [1] the Nissen dataset (72 stars, top row), [2] stars that are in common between the  Nissen dataset  and ours (52 stars, second row), [3] the dataset under investigation in this work (485 stars, third row), [4] the dataset from \citet{Miglio_2021} (2785 stars, bottom row; typical age error   1 Gyr).

First, we analyzed Nissen's dataset with the method described in Sect. \ref{method}. We took into account the various sources of uncertainty (Sect. \ref{method}), and we can see that the analysis shows that the  separation between the two populations is not relevant. The drop in the star's distribution across the $d_z$ axis is less than 1$\sigma$ uncertainty (Fig. \ref{fig: AMR}, top row). All of this suggests that the separation found by \citet{Nissen_2020} is due to the small number of statistics.

Second, our dataset contains  52 stars in common with \citet{Nissen_2020}, and  can be seen in Fig. \ref{fig:age comparison}, right, the metallicity measurements are in perfect agreement with \citet{Nissen_2020}. In addition, as we discuss in Sect. \ref{age determination}, the ages derived in this work and in \citet{Nissen_2020} have the same precision and are in good agreement with each other. By applying the method for the overlapping 52 stars with age and metallicity derived by us we can   see the same pattern as in \citet{Nissen_2020} data; however,   when taking into account all the uncertainties, the separation between the two populations is not as evident as in the  \citet{Nissen_2020} data (Fig. \ref{fig: AMR}, second row). The analyses with different ages and metallicity determinations both show the same pattern, but we can conclude that taking into account uncertainties the \citet{Nissen_2020} dataset does not contain enough stars to support a strong claim about the presence of two populations.

Third, we analyzed the whole sample from this work. In Fig. \ref{fig: AMR}, third row, we do not see any signature of the two populations. The resulting distribution is in good agreement within the uncertainties with the model of homogeneously distributed stars in the age-metallicity map (Fig. \ref{fig: Method}, bottom right). Moreover, we do not detect any double structure in the age-metallicity plot (Fig. \ref{fig: AMR}, third-row middle) and in the age-metallicity density map (Fig. \ref{fig: AMR}, third row left).

In the end, we also analyzed the data from \citet{Miglio_2021} via our method. Their dataset is composed of red giant and red clump stars with APOGEE metallicity measurements and asteroseismic high-accuracy ages (average uncertainty: 1 Gyr). We cut their dataset in metallicity to match our range of study ($-0.3 < [Fe/H] < 0.3 dex$). In Fig. \ref{fig: AMR}, bottom right, we detect a single population centered around 3 Gyr with smoothly decreasing density toward older ages. This effect can be explained by the fact that the selected stars are from the red giant branch and red clump region. Stars spend less time in the red giant branch and the red clump phases than in the main sequence. Therefore, these two regions are mainly populated by stars with a specific narrow mass range, and  hence ages. However, for the main sequence this range of masses is bigger, and as a result we have a more horizontal distribution in age. That is why we do not see the homogeneous distribution across the $d_z$ axis as we see for our dataset of solar-twin stars. However, for the \citet{Miglio_2021} dataset we also do not detect the presence of the two populations found by \citet{Nissen_2020}.

\subsubsection{Unaccounted age uncertainties}
\begin{figure}
\centering
\includegraphics[scale=0.5]{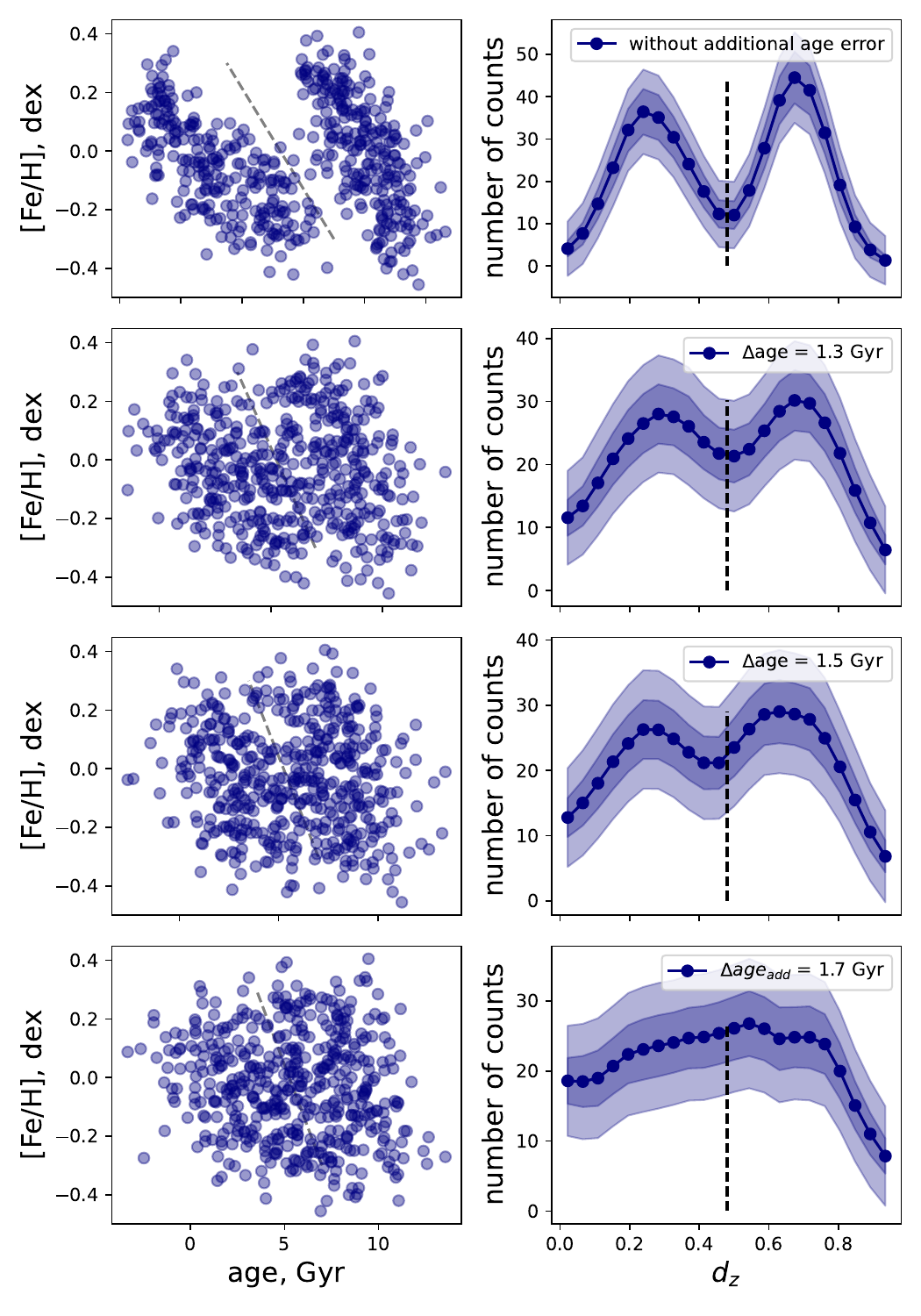}
\caption{Test for the ability of unaccounted uncertainties in age determination method to eliminate the presence of two populations. In each line, the results of distinguishing two populations are presented with unaccounted errors: 0.0, 1.3, 1.5, 1.7 Gyr from top to bottom. We need at least 1.7 Gyr of additional unaccounted error in age to eliminate any signature of double populations.}
\label{fig: extra error test}
\end{figure}

It may still be argued that we have unaccounted age uncertainties that blur over the separation. Therefore, we performed a test to estimate the value of unaccounted uncertainty we need to blur over the separation between two populations similar to those of \citet{Nissen_2020}. To do that, we simulated the dataset of 485 stars (the same number as the dataset under investigation) with two separate populations in the exact locations used in \citet{Nissen_2020} (Fig. \ref{fig: extra error test}, first row). We took three main points from younger ((age$_{[Gyr]}$, [Fe/H]$_{[dex]}$): (7, 0.2), (8, 0), (9, -0.2)) and from older ((age$_{[Gyr]}$, [Fe/H]$_{[dex]}$): (1.5, 0.1), (3, -0.1), (5, -0.2)) populations, and randomly created blobs with Gaussian distribution with sigma equal 0.7 Gyr for age and 0.1 dex for metallicity (Fig. \ref{fig: extra error test}, top left). For the measurement uncertainties, we used 0.06 dex for metallicity and 0.7 Gyr for age, which are the average uncertainties for our dataset. As a result, we see in Fig. \ref{fig: extra error test}, top right, that the drop in density distribution has a $2\sigma$ significance. Then for each artificial star, we randomly selected an age value inside the Gaussian distribution with the sigma equal to an unaccounted error. This approach simulates the effect of blurring due to unaccounted error in age. As a result, in Fig. \ref{fig: extra error test} we can see that the drop due to the presence of two populations disappears when an unaccounted error is equal to 1.7 Gyr. This value should be understood as an possible unknown uncertainty that we have to have in addition to 0.7 Gyr age determination uncertainty to be able to explain the absence of two populations through the blurring effect. This value is 2.5 times bigger than the age determination uncertainty and also two times more than the direct error of age dispersion between our ages and those from \citet{Nissen_2020} (Sect. \ref{age determination}). It means that it is very unlikely that we have such a huge unaccounted error and, as a result, it is also unlikely that we have a significant blurring effect.

\subsubsection{Selection bias}
While studying the Milky Way formation and evolution it is very important to pay attention to the selection function effects. Cutoffs in color-magnitude diagrams, distance limits, and other selection criteria misrepresent the real picture of our Galaxy. As we have already discussed, for the \citet{Miglio_2021} dataset selecting stars from the red giant branch--red clump region (fast evolutionary phases) creates a peak in age at around 3 Gyr. \citet{Feuillet_2018}, however, found a double-peaked structure due to selection bias. In addition, as shown in \citet{Sahlholdt_2022}, a distance larger than 600 pc creates a bias in the age-metallicity map changing the initially homogeneous distribution to the distribution with several features. As a result, the selection function plays an important role in the study of the Milky Way formation and evolution. 

In our work, we selected stars from the main sequence--turn-off point as this phase is much slower compared to the red giant branch--red clump phase, and therefore our data should be much less contorted. Most of our stars are also located inside a 200 pc bubble, which is far below the limit found by \citet{Sahlholdt_2022}. Therefore, we can conclude that based on our knowledge our data is not affected by the selection bias and represents the original AMR in the solar neighborhood.

\begin{figure*}
\centering
\includegraphics[scale=0.5]{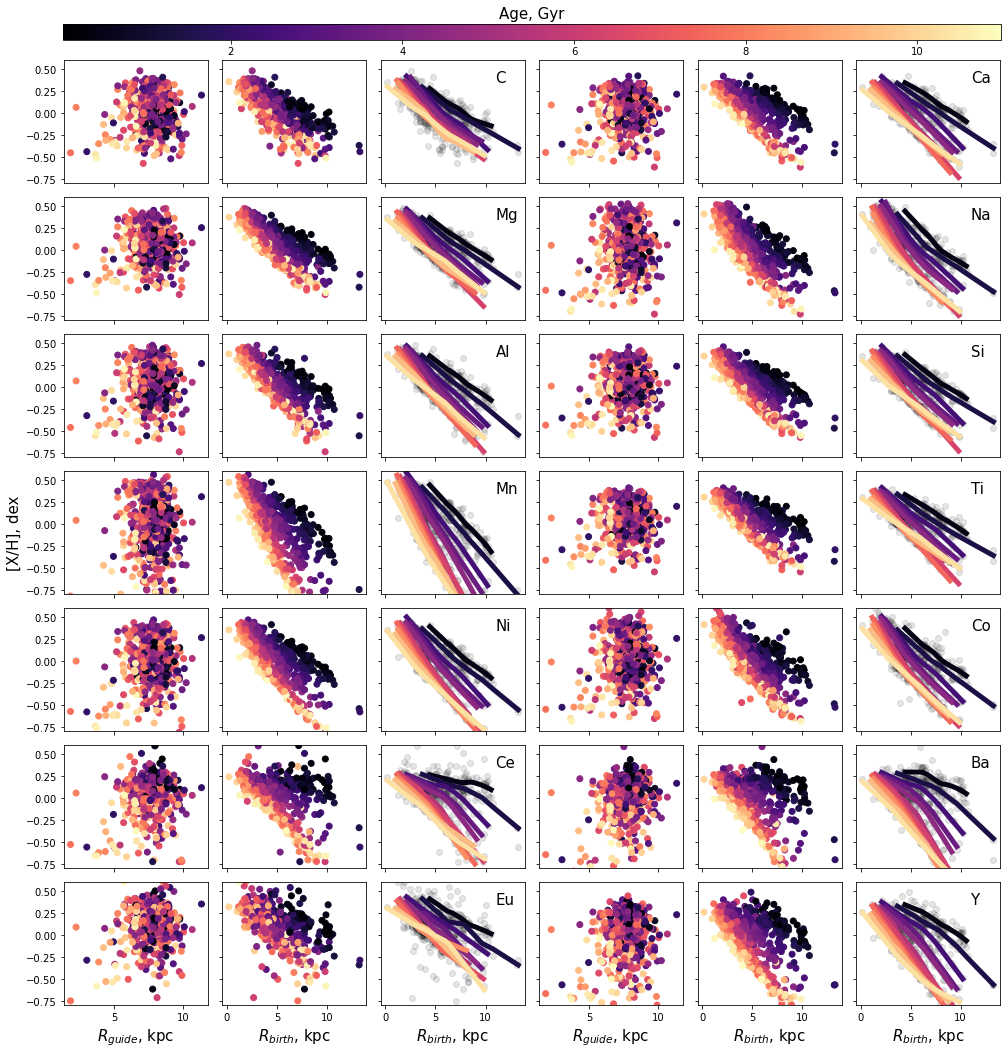}
\caption{Chemical trends for 14 elements with guiding and birth radii ($R_{guide}$, $R_{birth}$) color-coded by age. In the third and sixth columns the trend lines were obtained by locally weighted scatterplot smoothing (\texttt{LOWESS}).}
\label{fig: chemical trends}
\end{figure*}

\subsection{Age-metallicity relation summary}
To sum up, the separation seen for stars selected in the \citet{Nissen_2020} dataset is not significant if we consider all sources of uncertainty; the drop in star counts due to separation is less than $1\sigma$. An increased number of stars with the same high-quality high signal-to-noise metallicity measurements coupled with accurate age values also do not show any signature of the separation. This seems to indicate that the presence of two populations in the  \citet{Nissen_2020} study is probably due to a sort of selection bias coupled with very low statistics. The comparison of this result with other studies, as we discussed in Sect. \ref{method}, shows that only in the case of adopting ages from the APOGEE based on C and N dependencies \citep[][]{Jofre_2021} shows the presence of two populations. All the other studies do not show it (\citet{Xiang_2022}, nor do studies based on solar twins: \citet{Spina_2016}, \citet{Spina_2018}, \citet{Bedell_2018}, \citet{Casali_2020}). The study of \citet{Miglio_2021}, which uses APOGEE stellar parameters and abundances but Kepler asteroseismic ages, also does not show any separation (Fig. \ref{fig: AMR}, bottom row). This implies that the age determination method plays an important role,  and that APOGEE ages may have a sort of bias that creates a nonlinear distribution in age. For example, \citet{Anders_2023} showed that the residuals between their ages and those derived with other methods are not linear, but display an oscillation pattern. That could cause a nonlinear distribution of stars in the age-metallicity map. This hypothesis should be checked with a deeper study.  

\begin{figure}
\centering
\includegraphics[scale=0.39]{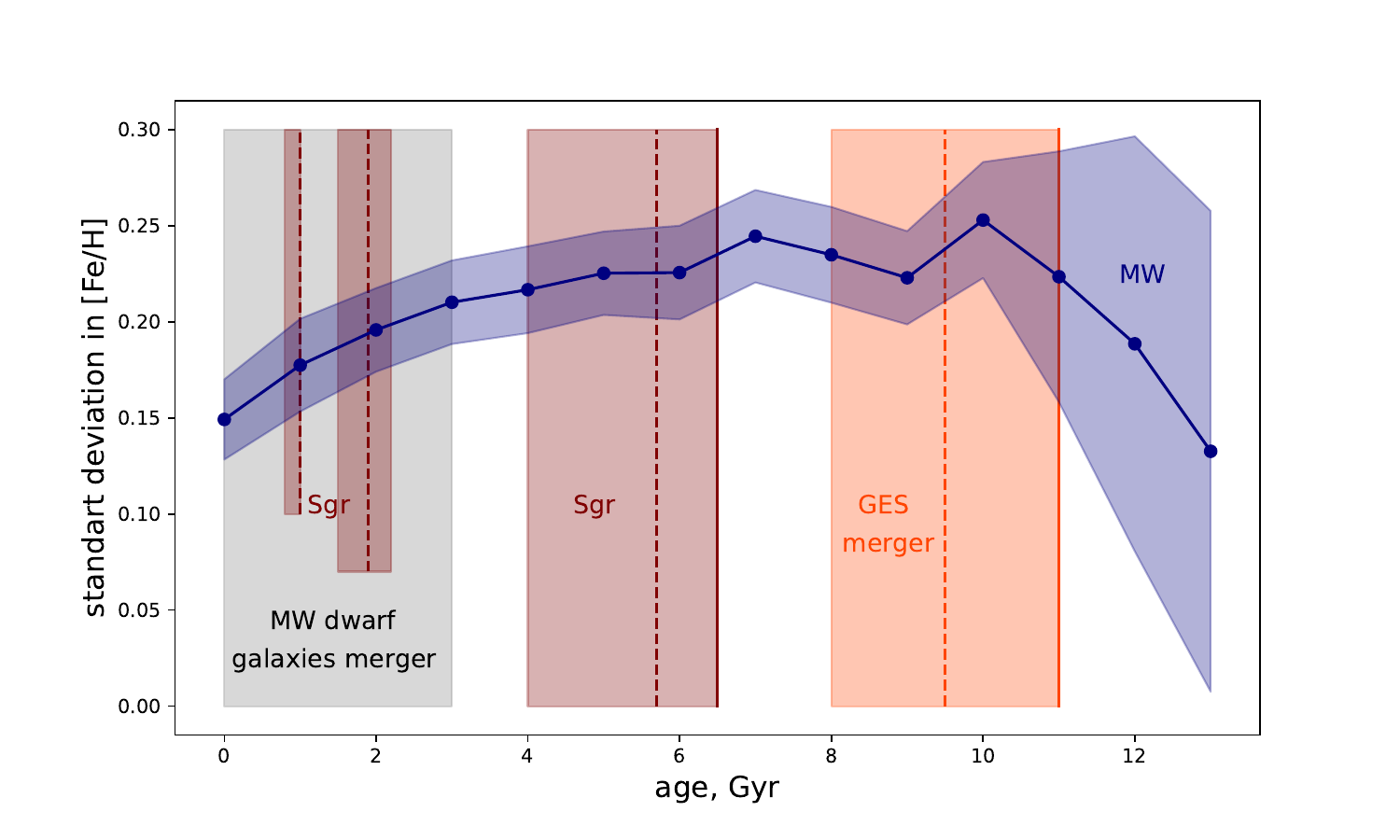}
\caption{Metallicity standard deviation vs. age trend of the stars under investigation (blue line for- main trend; blue region for  measurement uncertainty Sect. \ref{method}). Orange: GES accretion event (observations: \citet[][]{Ruiz-Lara_2020, Gondoin_2023, Lu_2022, Ratcliffe_2023, Anders_2023}, numerical simulations: \citet[][]{Laporte_2019, Belokurov2018, Helmi_2018, Buck_2023, Wang_2024_Sgt}); red: Sgr \citep{Ruiz-Lara_2020}; gray: mergers of the Milky Way dwarf galaxies \citep{Wang_2024_Sgt}. The solid vertical lines show the beginning of the merger, and the dashed vertical lines show the peak of star formation corresponding to a particular accretion event.}
\label{fig: std}
\end{figure}

\section{Chemical trends}\label{Chemical trends}
Chemical abundances, ages, metallicity, guiding radius, and birth radius all together allow the chemical evolution of the Milky Way disk to be traced.  Figure \ref{fig: chemical trends} illustrates chemical trends versus birth and guiding radius color-coded with age for 14 chemical elements. As for the guiding radius, we find that stars of different ages are well mixed, and this high dispersion makes it difficult to trace the chemical evolution with age. On the other hand, exploiting the birth radius trends, we see that stars are separated from each other. This implies that radial migration (the basis of the chemical evolution model in the birth radius determination method \citep[][]{Minchev_2018, Lu_2022, Ratcliffe_2023}) explains the basic age-chemistry dependences in the disk region around the Sun. However, we note the presence of knees and a slight overlap in the age ranges 4 - 6 Gyr and 8 - 12 Gyr (Fig. \ref{fig: chemical trends}). Therefore, some other mechanism should be at work in tandem with radial migration. The most plausible mechanism is surely the cumulative effect of the accretion that the Galaxy has experienced since its early assembly and that must have had an impact on the disk evolution in particular. This has been pointed out in several recent observational studies, which for instance illustrate the role of the GES accretion event in altering the chemical evolution of the Milky Way disk in the period from 8 to 11 Gyr \citep[][to give a few examples]{Ruiz-Lara_2020, Gondoin_2023, Lu_2022, Ratcliffe_2023, Anders_2023}. These effects have also been reproduced by numerical simulations \citep{Laporte_2019, Belokurov2018, Helmi_2018, Buck_2023, Wang_2024_Sgt}. The period in which GES was effective is depicted in Fig. \ref{fig: std} (orange region). 

In addition, several studies have demonstrated the significant impact of the Sgr pericentric passages (first: 4 - 6.5 Gyr; second: 1.5 - 2.2 Gyr; third: 0.4 - 1 Gyr, Fig. \ref{fig: std}, red region, \citet{Ruiz-Lara_2020}) in the history of the Milky Way disk formation around the Sun. Finally, \citet{Wang_2024_Sgt} showed that the Sgr passage affected the Milky Way in the past 3 Gyr,  and also other passages of Milky Way dwarf galaxies (Fig. \ref{fig: std}, gray region).  In addition, in Fig. \ref{fig: std} we present the metallicity dispersion versus age plot for our dataset, which shows the fluctuation corresponding to GES merger in the range from 8 to 10 Gyr. Additionally, we note a small fluctuation corresponding to the Sgr first pericentric passage around 5 to 6.5 Gyr, but this cannot be considered significant due to the uncertainty.

% \begin{figure}
% \centering
% \includegraphics[scale=0.35]{Plots/chemical trends.png}
% \caption{}
% \label{fig: chemical trends}
% \end{figure}

\section{Conclusions}\label{conclusion}
In this study we analyzed 114 solar twins for which we performed a spectroscopic analysis of high-resolution spectra from HARPS-N. We also analyzed a dataset from \citet{Casali_2020} with spectroscopic analysis from \citet{Casali_2020} (371 stars). For both datasets, we derived ages with the isochrone fitting technique in the effective temperature versus logarithm of surface gravity space with a metallicity correction for the atomic diffusion at the solar age. As a result, we obtained high-precision ages and chemical abundances that allowed us to study the AMR for solar twins around the Sun ([Fe/H]: -0.3 - 0.3 dex). We used a new parameter ($d_z$; Sect. \ref{method}) to test the AMR separation into two groups for solar twins in the solar vicinity. In the process, we took into account all possible sources of errors and did not detect a separation (Fig. \ref{fig: AMR}). This result is in agreement with \citet{Miglio_2021}, \citet{Xiang_2022}, and \citet{Lu_2022}. For \citet{Nissen_2020}, the separation was caused by the statistical bias. In the literature, there are studies that reached the opposite result \citep[][]{Jofre_2021, Ratcliffe_2023}. There are two possible reasons why these two studies show a separation. First of all, they employed ages obtained from the use of C and N abundances. This indicates that age determination is one of the crucial aspects of studying these chemical trends. Another extremely important parameter that   should always be tested while studying data that has cutoffs is the selection bias. Cutoffs in distance, magnitude, metallicity,  for example,  produce features that lead to misinterpretation of the data. This topic should be widely explored.

We also studied the dependence of chemical abundance with guiding and birth radii and age (Fig. \ref{fig: chemical trends}). We see that mixed star populations for the guiding radius transform into well-separated trends for the birth radius. This means that radial migration, which is the basis of the birth radii determination, explains well the star's chemical distribution around the Sun, but some effects, such as  the presence of knees for some age trends of some elements or overlap for some age trends, lead us to the conclusion that radial migration alone does not explain all the features in the Milky Way formation history.

Several authors have already showed that GES mergers and Sgr pericentric passages, as well as other passages and accretions of dwarf galaxies, have affected the Milky Way evolution. In Fig. \ref{fig: std} we also show fluctuations in the standard deviation of metallicity versus age for the GES merger (8 - 10 Gyr) and for the first pericentric passage of Sgr (5 - 6.5 Gyr). However, it is worth mentioning that taking into account uncertainties is a very important aspect of the statistical study. The fluctuations we detected for GES and Sgr are not significant taking into account the uncertainty.

%-------------------------------------------------------------------

% \section{Conclusions}

%    \begin{enumerate}
%       \item The conditions for the stability of static, radiative
%          layers in gas spheres, as described by Baker's (\citeyear{baker})
%          standard one-zone model, can be expressed as stability
%          equations of state. These stability equations of state depend
%          only on the local thermodynamic state of the layer.
%       \item If the constitutive relations -- equations of state and
%          Rosseland mean opacities -- are specified, the stability
%          equations of state can be evaluated without specifying
%          properties of the layer.
%       \item For solar composition gas the $\kappa$-mechanism is
%          working in the regions of the ice and dust features
%          in the opacities, the $\mathrm{H}_2$ dissociation and the
%          combined H, first He ionization zone, as
%          indicated by vibrational instability. These regions
%          of instability are much larger in extent and degree of
%          instability than the second He ionization zone
%          that drives the Cephe{\"\i}d pulsations.
%    \end{enumerate}

\begin{acknowledgements}
      The comments of an anonymous referee have been much appreciated. AP acknowledges Roman Tkachenko for useful consultations. BR acknowledges support by the Deutsche Forschungsgemeinschaft under the grant MI 2009/2-1. 
\end{acknowledgements}

% The comments of an anonymous referee have been much appreciated.

  \bibliographystyle{aa} % style aa.bst
  \bibliography{bibliography} % your references Yourfile.bib

% \begin{thebibliography}{}

%   \bibitem[Baker(1966)]{baker} Baker, N. 1966,
%       in Stellar Evolution,
%       ed.\ R. F. Stein,\& A. G. W. Cameron
%       (Plenum, New York) 333

%    \bibitem[Balluch(1988)]{balluch} Balluch, M. 1988,
%       A\&A, 200, 58

%    \bibitem[Cox(1980)]{cox} Cox, J. P. 1980,
%       Theory of Stellar Pulsation
%       (Princeton University Press, Princeton) 165

%    \bibitem[Cox(1969)]{cox69} Cox, A. N.,\& Stewart, J. N. 1969,
%       Academia Nauk, Scientific Information 15, 1

%    \bibitem[Mizuno(1980)]{mizuno} Mizuno H. 1980,
%       Prog. Theor. Phys., 64, 544
   
%    \bibitem[Tscharnuter(1987)]{tscharnuter} Tscharnuter W. M. 1987,
%       A\&A, 188, 55
  
%    \bibitem[Terlevich(1992)]{terlevich} Terlevich, R. 1992, in ASP Conf. Ser. 31, 
%       Relationships between Active Galactic Nuclei and Starburst Galaxies, 
%       ed. A. V. Filippenko, 13

%    \bibitem[Yorke(1980a)]{yorke80a} Yorke, H. W. 1980a,
%       A\&A, 86, 286

%    \bibitem[Zheng(1997)]{zheng} Zheng, W., Davidsen, A. F., Tytler, D. \& Kriss, G. A.
%       1997, preprint
% \end{thebibliography}

\end{document}